# Analysis of long range order

T.R.S. Prasanna

Department of Metallurgical Engineering and Materials Science

Indian Institute of Technology, Bombay

Mumbai, 400076 India

A first principles analysis of order-disorder transition in alloys shows that ordering energy is a function of temperature due to thermal vibrations. The inter-nuclear potential energy term converges if zero point vibrations are incorporated and this method can replace the Ewald sum method. Core energy contributions to the ordering energy are stored exclusively in superlattice lines. The effect of electron-phonon interactions on ordering energy is of the *same order of magnitude* as ordering energy near transition temperatures and cannot be ignored. Ising model and variants are incorrect in explaining alloy phase transitions as they ignore the role of electron-phonon interactions without justification. A theoretical formalism that incorporates the Debye-Waller Factor component of electron-phonon interactions in electronic structure calculations already exists and must be adopted when modeling temperature dependent phenomena. It is suggested that DWF correction will account substantially for the discrepancy between experimental and theoretical ordering energy in $Ni_3V$. Thermal vibrations alter magnetic ordering energy at finite temperatures. The role of electron-phonon interactions in alloy and magnetic phase transitions cannot be ignored and must be incorporated in all models. This will also ensure consistency with x-ray and electron diffraction (alloy transitions) and neutron diffraction (magnetic transitions) results. An isotope effect is predicted for (magnetic) phase transitions if the transition temperature is below Debye temperature. Recent observations of an isotope effect in magnetic phase transitions confirm our above conclusions and imply that the role of electron-phonon interactions must be incorporated in all theories and models of magnetism to avoid contradictions.

Order-disorder transitions have been the subject of intense study over decades. The first model for order-disorder transition in alloys was the Bragg-Williams (BW) model [1-3] that proposed the existence of a long-range order parameter. The deficiencies [1-3] of this model include a higher transition temperature, $T_c$, and failure to account for specific heat above $T_c$ due to the neglect of fluctuations. The cluster variation method [4] is a generalization of this approach that can be applied to the generalized Ising model where next nearest neighbor interactions are also incorporated. In parallel, models were proposed for magnetic ordering, the most influential among them being the Ising model. The Ising model is the simplest of nearest neighbor models that consider the cooperative interaction between neighboring spins. The exact solution to the two-dimensional (2D) Ising model [5] continues to be a landmark as the solution for the 3D Ising model has proved elusive. Recently, the author has described a method to determine the form of the exact partition function for the Ising model [6]. The BW model is equivalent to a mean-field solution of the Ising model.

The study of phase transitions has largely relied on models as first principles theories have proved difficult, especially in the study of magnetism. The prediction of phase diagrams [7-10] from first principles is another problem of great interest and one where considerable progress has been made. The development of density functional theory [11-13] is a landmark that has contributed significantly to this effort.

In this paper, a first principles analysis of order-disorder transition in alloys is presented which shows that ordering energy is temperature dependent due to thermal vibrations.

Some components of the total ordering energy are stored exclusively in superlattice vectors. Electron-phonon interactions shift electron energies by magnitudes that are of the same order of magnitude as ordering energy and hence cannot be ignored. A theoretical formalism that incorporates the Debye-Waller Factor component of electron-phonon interactions in electronic structure calculations *already exists* and must be adopted when modeling temperature dependent phenomena. Thermal vibrations alter the magnetic ordering energy at finite temperatures. Theoretical models, Ising, Heisenberg etc., are deficient as they do not incorporate the temperature dependence of ordering energy. All theoretical models must incorporate the role of thermal vibrations to be consistent with x-ray, electron and neutron diffraction results. Recent observations of an isotope effect in magnetic phase transitions confirm our conclusion that the role of electron-phonon interactions must be incorporated in all theories and models of magnetism.

In x-ray diffraction, the structure factor, $F_G$, is defined as the Fourier transform of the thermal-average charge density, $\overline{\rho(\mathbf{r})}$, and given by [14]

$$F_G = \int \overline{\rho(\mathbf{r})} e^{2\pi i \mathbf{G}\cdot\mathbf{r}} d^3\mathbf{r} \qquad (1)$$

and the intensity of a Bragg reflection is proportional to $|F_G|^2$. The intensity measured in a x-ray diffraction experiment is due to elastic scattering from time varying electron distributions that are in instantaneous equilibrium with various atomic configurations due to thermal vibrations [14,15]. However, for crystals the difference between the two intensities is negligible and $\Delta I/I_{Bragg} \sim O(N^{-1})$ [14]. Therefore, *"To an extremely good approximation, the scattering averaged over the instantaneous distributions is equivalent to the scattering of the time-averaged distribution of the scattering matter"* [16]. Hence, scattering of x-rays by crystals can be interpreted as resulting from a thermal-average electron density, $\overline{\rho(r)}$, in a static lattice [14-16]. At finite temperatures, thermal vibrations cause the thermal-average charge density to be diffused. In the harmonic approximation, with the assumption of rigid ions or pseudo atoms, the effect of thermal vibrations is to change the Fourier Transform of the atomic charge density, atomic scattering factor, $f_j$ to $f_j e^{-M_j}$ where $M_j$ is the Debye-Waller factor (DWF). [15-18]. Similar arguments are applicable in analysis of electron diffraction data [19].

The density functional theorem [11-13] states that for a given external potential V(**r**)**,** the ground state energy is a functional of the charge density, E[$\rho(r)$], and is minimized for the correct charge density. In solids, knowing the correct charge density, $\rho(r)$, is equivalent to knowing the correct structure factors, $F_G$, because of the uniqueness of Fourier Transforms. In principle, density functional theorem can be interpreted to mean that ground state energy is minimized for correct structure factors and can be represented as $E(F_G)$ though the exact functional dependence is unknown. In alloys, the energy of the disordered state can be represented as $E^{dis}(F_{G_f})$ because only fundamental lines are present. In the ordered state, superlattice lines are also present in addition to the fundamental lines and the energy can be represented as $E^{ord}(F_{G_f}, F_{G_s})$. The ordering energy can be symbolically represented as $\Delta E^{ord} = E^{ord}(F_{G_f}, F_{G_s}) - E^{dis}(F_{G_f})$. In the above discussion, it has been implicitly assumed that total energy is calculated at 0 K with no role for thermal vibrations. However, as discussed above, thermal vibrations alter the structure factors and the total ordering energy is more correctly represented as

$$\Delta E^{ord}(T) = E^{ord}(F_{G_f} e^{-M_{G_f}}, F_{G_s} e^{-M_{G_s}}) - E^{dis}(F_{G_f} e^{-M_{G_f}}) \qquad (2)$$

and is a function of temperature due to the role of thermal vibrations.

The above conclusion that ordering energy is a function of temperature is also supported by the following physical argument. At finite temperatures, various atomic configurations are accessed due to thermal vibrations and the electron distribution is in instantaneous equilibrium with them. The electron distributions in equilibrium with various configurations are of higher energy compared to the static lattice value. Clearly, the ensemble average over these configurations will result in higher energy when compared to the 0 K value. This result is valid for both the ordered and disordered state. At 0 K, the difference in energy between them is the conventional ordering energy. At any finite temperature, ordering energy is the difference in the ensemble average energies of the ordered and disordered state. It would be an extraordinary coincidence if this difference were to be *exactly* equal to the difference at 0 K and cannot be assumed without justification. Therefore, the correct conclusion is that the differences between the ordered and disordered state ensemble average energies at finite temperatures are different from the 0 K value. Hence, *ordering energy is a function of temperature*.

It has been observed experimentally that thermal vibrations shift electron energies by ~ 0.2 meV/K [30] in metals and semiconductors. This implies shifts in electron energies of ~ $10^2$ meV near transition temperatures (~ 1000 K) for both ordered and disordered phases. *Their difference would be of the same order of magnitude as the ordering energy (10-100 meV) [8] and hence, the role of electron-phonon interactions cannot be ignored.*

A theoretical formalism that incorporates the role of thermal vibrations through the DWF in electronic structure calculations already exists. These issues are discussed later.

In the above discussion, we have presented arguments that clearly show that ordering energy is a function of temperature. Below, we consider changes on various components of the ordering energy which will also provide an estimate of the significance of the role of thermal vibrations. The total energy can be represented as

$$E_{tot} = E_{kin} + E_{n-n} + E_{c-c} + E_{c-n} + E_{v-n} + E_{v-c} + E_{v-v} + E_{xc} \tag{3}$$

While they are usually evaluated in real space, Fourier (or reciprocal space) methods have also been developed [20,21]. The latter incorporate the periodicity of the lattice and various terms are expressed as sums over reciprocal lattice vectors [20,21].

We begin with an analysis of the nuclear-nuclear energy term. From standard textbooks in electron microscopy [19], the structure factor due to nuclear charges, $Z_G$, is given by

$$Z_G = \sum_{j=1}^{N} Z_j e^{-M_j} e^{2\pi i \mathbf{G} \cdot \mathbf{r}_j} \tag{4}$$

where $Z_j$ and $M_j$ are the charge and DWF for nucleus $j$, $\mathbf{r}_j$ its position in the unit cell and the summation is over all (N) atoms in the unit cell. **G** is a reciprocal lattice vector whose magnitude is given by $|\mathbf{G}|^2 = 1/d_{hkl}^2$ where $d_{hkl}$ is the interplanar spacing [18]. In the above expression, since the nuclear charge distribution can be considered to be 'rigid' with respect to thermal vibrations, DWF accounts for their role to a very good approximation. From the Poisson equation it follows that the Fourier transform of the potential due to nuclear charges is given by $V_G^n = Z_G / G^2$, *which is temperature* *dependent* and similar to those found in electron microscopy textbooks [19]. The nuclear-nuclear repulsion energy, $E_{n\text{-}n}$, is given by

$$E_{n-n} = \sum_{G \neq 0} \frac{|Z_G|^2 e^{-2M}}{G^2} \tag{5}$$

The same DWF (M) has been assumed for all atoms in the unit cell in Eq.5. The **G** = 0 term is not considered as the average charge and potential in a unit cell is zero [23].

Before, we analyze the change in $E_{n\text{-}n}$ due to ordering, we discuss the Ewald summation method of evaluating $E_{n\text{-}n}$. It is well known [22,23] that the expression for $E_{n\text{-}n}$ does not converge e.g. Eq.F13 of Ref.23 (appendix F, p-640) and artificial parameters are introduced to ensure convergence. This is because this term is usually represented in the static approximation which ignores the role of thermal vibrations. However, even at 0 K, zero point vibrations are present and their role must be incorporated. *Hence, Eq.5 is the correct representation of $E_{n\text{-}n}$*. It is readily seen that with M = 0, Eq.5 reduces to Eq.F13 of Ref.23. The low temperature approximation to DWF, $M_L$, must be used in Eq.5 and is given by [23] $M_L = 3h^2|G|^2/8mk\Theta_D$ where $|G|^2 = 1/d_{hkl}^2$. It is readily seen that Eq.5 converges due to the presence of an exponential term that decays as $e^{-|G|^2}$ ignoring constants. Thus the problem of convergence of $E_{n\text{-}n}$ is overcome if zero point vibrations are incorporated. Since Eq.5 represents the physics correctly and does not require any artificial parameters, it should replace the Ewald sum technique in evaluating $E_{n\text{-}n}$. An important advantage of this method is that it allows $E_{n\text{-}n}$ to be readily evaluated at any temperature.

We next consider the change in the nuclear-nuclear energy term upon ordering. There is no change in structure factors, $Z_G$, upon ordering as the nuclear charge distribution remains "frozen". In the disordered phase, the summation in Eq.5 is only over fundamental lines, $G_f$. In the ordered phase, the summation in Eq.5 is over fundamental lines, $G_f$, and superlattice lines, $G_s$. Comparison of the two shows that the change in nuclear-nuclear energy upon ordering, $\Delta E_{n-n}^{ord}$, *is stored only in superlattice wavevectors* and is given by

$$\Delta E_{n-n}^{ord} = \sum_{G_s} \frac{|Z_G|^2 e^{-2M}}{G^2} \tag{6}$$

At finite temperatures, the high temperature form, $M_H$, given by [18,23]

$$M_H = \frac{3h^2 |G|^2 T}{2mk\Theta_D^2} \tag{7}$$

must be used. We stress that in both Eq.5 and Eq.6, we have approximated the DWF for different atoms, represented as $M_j$ in Eq.4, to be the same ($M$). This has been done to highlight explicitly the role of thermal vibrations on the $E_{n-n}$ energy term. More correctly, this energy term is represented as $E_{n-n} = |Z_G|^2 / G^2$ where $Z_G$ is given by Eq.4. It is readily seen that even in this more correct form, $E_{n-n}$ will converge due to the $M_j$ term.

We consider next the electron-electron energy term, $E_{e-e}$, given in DFT by [11-13]

$$E_{e-e} = \frac{1}{2} \iint \frac{\rho(r)\rho(r')}{|r-r'|} d^3r\, d^3r' \tag{8}$$

where $\rho(r)$ is the total charge density. At finite temperatures, thermal vibrations result in various configurations being accessed, but to a very good approximation [14-16] charge density can be represented in terms of the thermal-average, $\overline{\rho(r)}$, as discussed earlier. The core electron density is in principle a function of the chemical environment but is also frequently assumed to be 'rigid'. In this case, the effect of thermal vibrations can be incorporated by a DWF correction to the structure factor. The core electron – core electron interaction energy, $E_{c-c}$, is given in terms of Fourier components by

$$E_{c-c} = \sum_{G \neq 0} \frac{|F_G^c|^2 e^{-2M}}{G^2} \tag{9}$$

assuming same DWF (M) for all atoms. $F_G^c$ is also given by

$$F_G^c = \sum_{j=1}^{N} f_j^c e^{-M_j} e^{2\pi i G \cdot r_j} \tag{10}$$

where $f_j^c$ is the atomic scattering factor of core electrons and is the Fourier Transform of the core atomic charge density of atom $j$. The potential due to core electrons is $V_G^c = F_G^c / G^2$ and *is temperature dependent*. Following arguments as in case of $E_{n-n}$, the entire change in core electron-electron energy upon ordering, $\Delta E_{c-c}^{ord}$, is given by

$$\Delta E_{c-c}^{ord} = \sum_{G_s} \frac{|F_G^c|^2 e^{-2M}}{G^2} \tag{11}$$

and *stored only in superlattice lines*. We also see from Eq.11 that the energy is a function of temperature due to the DWF factor. From the above discussion, it follows that the ordering energy between the nucleus and core electrons, $\Delta E_{n-c}^{ord}$, is also a function of temperature of the type $e^{-2M}$ and *stored only in superlattice lines*. Therefore, all core energy components of the total ordering energy, $\Delta E_{n-n}^{ord}$, $\Delta E_{c-c}^{ord}$, $\Delta E_{n-c}^{ord}$, *are stored only* *in superlattice lines* and are functions of temperature. All theoretical models, including cluster expansion models [10] must account for these experimentally derived facts.

The above analysis in based on experimental observations that Bragg intensities in x-ray and electron diffraction are temperature dependent, which implies that structure factors are temperature dependent [18,19]. It follows that the core lattice potential is temperature dependent and is most simply represented by $V_G^n = Z_G/\boldsymbol{G}^2$ and $V_G^c = F_G^c/\boldsymbol{G}^2$, where $Z_G$ and $F_G^c$ are given by Eq.4 and Eq.10. This potential incorporates the role of DWF and represents the average potential at high temperatures. Since the core lattice potential is different at high temperatures from the static lattice (0 K) potential, it follows that the valence electron distribution and energies will also be different. *Therefore, analysis of Bragg intensities shows that both core and valence electron components of the ordering energy are temperature dependent*. It is readily seen that $\Delta E_{v-n}^{ord}$ and $\Delta E_{v-c}^{ord}$ will contain *at least* a temperature dependence of the type $e^{-M}$ that arises from the nuclear charge or core electron structure factors. It is difficult to analyze changes in kinetic energy, valence electron-electron energy, exchange and correlation energy terms upon ordering.

So far, we have not considered any magnitudes of the effect of thermal vibrations. We consider beta-brass, $\beta$-CuZn, which exhibits an order-disorder transition [24,25] at 741 K. It has a Debye temperature, $\Theta_D = 263$ K [26] and lattice parameter of a = 2.95 Å. Instead of calculating the DWF, 2M, for which temperature needs to be specified, we write as 2M = αT where the high temperature DWF (Eq.7) has been used. We see that α is independent of temperature. Its values for different superlattice lines are $α_{100} = 1.5 \; 10^{-4}$ K⁻

[1] for (100), $\alpha_{111}$ = 4.5 $10^{-4}$ $K^{-1}$ for (111) and $\alpha_{300}$ = 1.3 $10^{-3}$ $K^{-1}$ for superlattice line (300). Usually, total energy is evaluated for a static lattice (M = 0) in electronic structure calculations. The change in $\Delta E_{n-n}^{ord}$ at 725 K and 0 K (static lattice) can be determined from Eq.5 and Eq.7. For $\beta$-CuZn, the inter-nuclear energy stored in different superlattice lines are reduced substantially. At 725 K, $\Delta E_{n-n}^{ord}(100)$, $\Delta E_{n-n}^{ord}(111)$ and $\Delta E_{n-n}^{ord}(300)$ are reduced to 89%, 72% and 37% of their respective 0 K values. Similar reductions would also occur for $\Delta E_{c-c}^{ord}$ and $\Delta E_{n-c}^{ord}$ as they have identical temperature dependence. These large changes in core energy contributions are alone sufficient to show that the Ising model and its variants (higher nearest neighbors etc.) are incorrect as they do not contain a temperature dependence of ordering energy. Also, the DWF are different in the ordered and disorder phases of $\beta$-CuZn [24,25] which will contribute additionally to the difference in ordering energy at high temperatures compared to the 0 K value.

The simplest model of phase transitions is the Bragg-Williams model and the ordering energy can be made temperature dependent in the Modified Bragg-Williams model as

$$-\Delta E^{ord}(T) = E_0 e^{-\alpha_m T} \tag{12}$$

where $E_0$ is the total disordering energy at 0 K in the BW model. The configuration entropy remains same in BW and MBW models with the assumption of random distribution. The critical temperature, $T_c$, is given by

$$T_c = \frac{2 E_0 e^{-\alpha_m T_c}}{R} = T_c^{mft} e^{-\alpha_m T_c} \tag{13}$$

In BW model, the (mean field) critical temperature is given by $RT_c^{mft} = 2E_0$. The value of parameter $\alpha_m$ must be representative of thermal vibrations ($\sim 10^{-3}$-$10^{-4}$ K$^{-1}$).

For beta-brass, β-CuZn, the experimental T$_c$ is 741 K. Assuming a nearest neighbor Ising model [24,25] with 8 neighbors, the mean field (BW) critical temperature is given by the relation $T_c/T_c^{mft} = 0.79385$ [27] to be $T_c^{mft} = 933.4$ K. Assuming that T$_c$ in MBW model is the experimental value, 741 K, gives $\alpha_m$ to be 3.11 10$^{-4}$ K$^{-1}$. This is within the range set by the lowest and highest observed superlattice lines as $\alpha_{100}$ is 1.5 10$^{-4}$ K$^{-1}$ and $\alpha_{300}$ is 1.3 10$^{-3}$ K$^{-1}$ as discussed above. Thus a mean $\alpha_m$ that is representative of thermal vibrations is sufficient to lower T$_c$ from the mean-field value of 933 K to the observed value of 741 K. It is stressed that the above discussion is meant to highlight the role of thermal vibrations and is not meant to be accurate. This is seen from the expression for the ordering energy, $E_0 e^{-\alpha_m T}$, where the exponential form has been adapted from the inter-nuclear and core electron energy terms, Eq.6 and Eq.11, as the temperature dependence of other energy terms is unknown. However, it is readily seen that T$_c$ will be altered substantially if even a fraction of the total ordering energy has exponential temperature dependence. Thus, even a simple MBW model shows that the role of thermal vibrations cannot be ignored.

The main focus of first principles theories or models has been to compare theoretically obtained order parameter η vs T behavior, T$_c$ and critical behavior with experimental observations. They completely ignore the need to account for the temperature dependence of Bragg intensities of superlattice lines that is routinely observed in x-ray and electron diffraction. If experimentalists were to ignore thermal vibrations in analysis of diffraction data to be consistent with theoretical models, it would lead to absurd consequences that different superlattice lines give different order parameters. Since the role of thermal vibrations is an intrinsic part of analysis of experimental diffraction data, it follows that theoretical models must do the same. Hence, the temperature dependence of Bragg intensities of superlattice lines is *an additional criterion* that must be explained by all theories of order-disorder phase transitions.

As discussed earlier, temperature dependence of Bragg intensities suggests that the core lattice potential must be represented as $V_G^n = Z_G/\boldsymbol{G}^2$ and $V_G^c = F_G^c/\boldsymbol{G}^2$ where $Z_G$ and $F_G^c$ are given by Eq.4 and Eq.10. These temperature dependent potentials are also justified by electron channeling experiments [28,29]. Such a theoretical formalism *already exists* [30-37] and it incorporates the DWF component of electron-phonon interactions in electronic structure calculations. Ref.32-37 and their citations clearly show that this formalism the first recourse to explain the temperature dependence of valence electron properties in metals and semiconductors and hence must also be used to determine the temperature dependence of ordering energy.

In addition to changes in lattice parameters, thermal vibrations result in electron-phonon interactions that lead to DWF and self-energy contributions at high temperatures [30,31]. These are universal effects in all solids at finite temperatures [30,31]. The correct approach to incorporate these effects has been described [30] as "*A higher order adiabatic perturbation summation can be accomplished by solving* $H_0 + \overline{H}_2 + \overline{H}_4 + ...$ *exactly (Keffer et. al. 1968) and then using the resulting temperature-dependent* *eigenfunctions and energies to calculate the self-energy terms*". That is, the first step is to incorporate the DWF in electronic structure calculations and the next step is to use the eigenfunctions and energies to calculate the self energy corrections. Therefore, even if self energy corrections were to be neglected (as in static lattice calculations), at the very least, the DWF correction [32-37] must be incorporated in modeling temperature dependent phenomena. This is equivalent to performing electronic structure calculations for a static lattice at high temperatures. Clearly, *this is a better approximation for modeling the high temperature state* than the current practice of performing static lattice calculations at 0 K. Another reason as to why *at least* the DWF correction must be incorporated is the different manner in which the DWF and self energy corrections affect the band structure. Self energy correction affects electrons within Debye energy of the Fermi level. While this is very important for properties that are strongly dependent on the behavior of electrons near the Fermi level, it does not affect all valence electrons, e.g. the density of states distribution is unlikely to be very significantly altered. On the other hand, the DWF correction affects the behavior of all valence electrons as they all move in the background of a temperature dependent potential. In particular, the density of states distribution is likely to be altered by DWF correction as is seen for Cd [34]. Since it is essential to know the density of states distribution for total energy calculations, *at the very least the DWF correction must be incorporated when modeling the high temperature state*. Hence, the above formalism [30-37] must be adopted for modeling of phase diagrams and high temperature phase transitions where determination of total energies of various phases at high temperatures and especially their differences is of great significance. For higher accuracy, self energy corrections must be incorporated using results of DWF corrections as suggested by Ref.30.

Ising-like models completely neglect the effect of electron-phonon interactions on ordering energy at high temperatures. However, they must be properly accounted for as they are universal phenomena that occur in all solids at finite temperatures. These effects shift electron energies by 0.2 meV/K [30] in metals and semiconductors. The shift in energies (from 0 K values) at transition temperatures (~ 1000 K) would be ~$10^2$ meV for both the ordered and disordered phases. *Their difference would be of the same order of magnitude as the ordering energy (10-100 meV) [8] and hence thermal effects cannot be ignored in models of alloy phase transitions*.

Molecular dynamics simulations [38] also reveal the crucial role of thermal vibrations in altering band structures at high temperatures. The lattice stability of bcc Mo over fcc Mo is reduced to 0.2 eV at 3200 K from 0.4 eV at 0K. This shows that the role of thermal vibrations on lattice stability is of the same order of magnitude as lattice stability itself. On the basis of the study of lattice stability [38], the authors speculate that the role of thermal vibrations in altering ordering energy in alloys at high temperatures also cannot be ignored which is identical to the our conclusion.

A general conclusion to the same effect, which anticipates the results of Ref.38, was reached by Allen and Hui [30] almost thirty years ago. They state "*The resulting additional terms in the free energy are not large compared to the harmonic oscillator* *background. However, suppose we wish to locate the temperature of a phase boundary. The harmonic term is not likely to differ too much between competing structures. The smaller electronic contribution may well be significant. The corrections we find have the same order of magnitude and thus may play a significant role*". This is very similar to the conclusion reached by us above that the effect of electron-phonon interactions on the ordering energy is of the same order of magnitude as the ordering energy itself and cannot be ignored. Clearly, the role of electron-phonon interactions on free energy becomes very significant whenever <u>difference in total energies</u> between two phases is to be considered, whether it is lattice stability, ordering energy or more generally, phase diagram calculations. However, the results of Allen and Hui [30], which are exact to second order perturbation theory, are not correct at high temperatures as the approximation $e^{-M}$ ~ 1-M (M is the DWF) is no longer valid as stated by Allen and Heine [30] and their suggested approach, perform electronic structure calculations with DWF corrections and use the results to calculate self energy corrections, must be adopted.

It has been suggested that since thermal vibrations smear the core lattice potential, valence electrons exhibit more free-electron like behavior [34] at higher temperatures. As this would be true of both the ordered and disordered phases, the difference in ordering energy at higher temperatures will be less than that at 0 K. Another argument leads to the same conclusion. The core lattice potential of ordered and disordered phases differ only in the contributions from superlattice wavevectors in the former. This includes a DWF and decreases at high temperatures due to thermal vibrations. Since the core lattice potentials are "closer" at high temperatures than at 0 K, it follows that the total energies will also be "closer" at high temperatures than at 0 K. This strongly suggests that ordering energy will be less at high temperatures than at 0 K.

The neglect of the effect of DWF component of electron-phonon interactions on ordering energy in Ising-like models is not deliberate but most probably due to lack of awareness. This is seen from the absence of Ref.30-37 in standard literature [7-9] on alloy theory. Conversely, none of the citations to Ref.30-37 pertain to alloy phase transitions. In particular, there is no mention of Allen and Hui's paper [30] even though they had mentioned in 1980 itself that corrections to the free energy from electron-phonon interactions are of the same order of magnitude as the difference in free energies between two competing structures and thus would be significant. Clearly, the role of electron-phonon interactions cannot be ignored in theoretical modeling of phase diagrams. The absence of Ref. 30-37 in standard literature on alloy theory [7-9] leads to the conclusion that the neglect of DWF corrections is most probably due to a lack of awareness of this phenomenon. It is clear that Ising-like models [3-10] are based on an unjustified assumption that ordering energy is unaffected by electron-phonon interactions. Only after it is shown that the ordering energy at 0 K and near $T_c$ are negligibly different using the formalism of Ref.30-37 can this assumption be justified. In the absence of such an exercise, any agreement of Ising-like models with experimental observations can only be a coincidence. Incorporating the role of electron-phonon interactions on ordering energy promises new insights into our understanding of phase transitions.

One example where the effect of electron-phonon interactions on ordering energy may be very significant is Ni$_3$V. In this alloy [39-41], the experimentally observed ordering energy (10 meV/atom) at 1400 K and the theoretically determined ordering energy (100 meV/atom) at 0 K differ by almost a factor of 10. Electronic excitations and spin polarizations [40,41] do not fully account for this discrepancy. Based on the above discussion, we see that the most probable reason for this discrepancy is the neglect of electron-phonon interactions that are present at 1400 K but absent in a static lattice calculation. Though the role of thermal vibrations on the discrepancy has been considered in Ref.40, it is <u>not</u> the DWF correction used in the formalism of Ref.30-37. The role of DWF on ordering energy can be estimated by performing electronic structure calculations at 0 K and at 1400 K using the above formalism [30-37]. This formalism is the *first recourse* to explain temperature dependence of valence electron properties in metals and semiconductors, as seen from Ref. 30-37 and citations to them. Therefore, there is no justification to ignore this formalism while accounting for the temperature dependent ordering energy in Ni$_3$V.

The results of the analysis of long range order can be summarized as follows. 1) It is an experimentally observed fact that Bragg intensities in x-ray and electron diffraction are temperature dependent. 2) This implies that the core lattice potential must be made temperature dependent by incorporating a DWF correction to its Fourier transform. 3) The core energy components of the total ordering energy, $\Delta E_{n-n}^{ord}$, $\Delta E_{c-c}^{ord}$ and $\Delta E_{n-c}^{ord}$, are functions of temperature with a simple form given by Eq.6 and Eq.11. 4) The core energy components of the total ordering energy, $\Delta E_{n-n}^{ord}$, $\Delta E_{c-c}^{ord}$ and $\Delta E_{n-c}^{ord}$, are stored exclusively in superlattice lines. 5) All valence energy terms will also be temperature dependent as electrons move in a temperature dependent core lattice potential. 6) Such formalism as proposed in 2) is well established in literature (Ref. 30-37 and citations) and is the *first recourse* to explain temperature dependence of valence electron properties in metals and semiconductors. Therefore, this formalism must be used whenever temperature dependent phenomena are to explained, including phase transitions. 7) This formalism incorporates the DWF component of (temperature dependent) electron-phonon interactions in electronic structure calculations. This is better than ignoring them altogether, especially when temperature dependent phenomena, e.g. phase transitions, are to be explained. 8) It is shown that the effect of electron-phonon interactions on ordering energy is of the same order of magnitude as ordering energy itself and hence cannot be ignored. 9) *Ising model and variants are incorrect as they ignore the role of electron-phonon interactions that result in a temperature dependent ordering energy.*

An important question is whether the conclusions from the analysis of alloy phase transitions are of sufficient generality to be valid for magnetic phase transitions as well.

Magnetic phase transitions are readily detected using neutron diffraction techniques [42-44] as ordering of spins gives rise to additional intensity either at new superlattice lines in case of anti-ferromagnetism or at fundamental lines in case of ferromagnetism. Magnetic scattering amplitudes and form factors are reduced by thermal vibrations at finite temperatures and are corrected by a Debye-Waller factor [42-44]. These experimental observations for magnetic ordering are identical to those in binary alloys where ordering is detected by x-ray diffraction. Therefore, the main conclusions from analysis of alloy ordering can be extended to magnetic phase transitions. This is also justified on phenomenological and theoretical grounds described below.

Most microscopic models (Ising, Heisenberg etc.) of magnetism [23,27,42-44] ignore the role of thermal vibrations thereby implicitly assuming a negligible role for them. According to this (implicit) model assumption, magnetic order parameter determined experimentally with or without taking into account the role of thermal vibrations should be negligibly different i.e. order parameter obtained from neutron diffraction data by including the Debye-Waller factor, M, or setting M equal to zero should be identical. However, if neutron diffraction intensities are not corrected for thermal vibrations, different superlattice lines will give different order parameter values. Therefore, correction for thermal vibrations is essential, which contradicts the implicit assumption of most models. The current situation is paradoxical where theoretical models ignore the role of thermal vibrations while experimentalists incorporate them in their analysis of experimental data. Since analysis of experimental data incorporates the role of thermal vibrations, theoretical models must do the same to be consistent with experiments.

In addition, the following argument also suggests a role of thermal vibrations in magnetic phase transitions. Magnetism originates in the exchange interaction involving the wavefunction of the unpaired electron and the total magnetic ordering energy depends crucially on the exchange energy. At finite temperatures, various nuclear configurations are accessed due to thermal vibrations and the electron distribution is in instantaneous equilibrium with them. Assuming that the exchange energy is independent of temperature is equivalent to assuming that the unpaired electron wavefunction is unchanged from the 0 K wavefunction for all configurations that are accessed due to thermal vibrations. It follows that the unpaired electron density and the magnetic form factor are identical at 0 K and finite temperatures. However, this is contradicted by neutron diffraction data where the magnetic form factor has a DWF dependence on temperature. This proves that the assumption that the unpaired electron wavefunction is unchanged by thermal vibrations at finite temperatures is incorrect. Therefore, it follows that the unpaired electron wavefunction is different for different configurations and the thermal-average exchange energy is a function of temperature.

As discussed earlier, electron-phonon interactions are universal phenomena that occur in all solids at finite temperatures. The strength of electron-phonon interactions is temperature dependent which implies that it must be accounted for especially when explaining temperature dependent phenomena, including magnetism. A simple formalism to incorporate the DWF component of electron-phonon interactions in electronic structure calculations already exists [30-37]. Incorporating the role of electron-phonon interactions through this formalism is much better than ignoring their role altogether (as is the current practice) in seeking an explanation of temperature dependent phenomena like magnetism. This formalism [30-37] results in a temperature dependent core lattice potential in whose background valence electrons move. It follows that the band structure of magnetic solids will be temperature dependent. In particular, electron wavefunction at finite temperatures would differ from the static lattice wavefunction. This would result in temperature dependent exchange and magnetic ordering energies. For a correct understanding of magnetism, this temperature dependence must be incorporated in all theories and models.

In the case of band ferromagnetism in Fe, Co and Ni, the density of states (DOS) at the Fermi level is a very important parameter [17]. Incorporation of the DWF alters the DOS [34] in Cd metal. MD simulations on Mo [38] also show that thermal vibrations alter the DOS at the Fermi level. Clearly, theories that seek to explain band ferromagnetism on the basis of DOS obtained from static lattice (0 K) calculations are incorrect. To correctly understand band ferromagnetism in Fe, Co and Ni, it is essential to incorporate the DWF in electronic structure calculations [30-37] to follow the changes in DOS at the Fermi level with temperature. This formalism will also account for the changes in exchange integral that will follow from changes in band structure at high temperatures.

Finally, we discuss the possibility of isotope effect in phase transitions due to the role of thermal vibrations. Thermal vibrations result in lattice expansions and also electron-phonon interactions that has two, DWF and self energy, contributions [30-37]. The DWF component of electron-phonon interactions can be easily incorporated in electronic structure calculations by a temperature dependent core lattice potential [30-37] as discussed earlier. Since DWF depends on mass, core lattice potentials will have isotopic dependence at finite temperatures. This will result in different valence electron wavefunctions and energies for different isotopes. Since the exchange energy depends crucially on the electron wavefunction, different isotopes will lead to different exchange energies at finite temperatures. This difference in energies will lead to an isotope effect on phase transitions. However, certain caveats are discussed below. For this purpose, it is better to consider the mean-square displacement that is related to the DWF of each ion, $M_j$ [18,45]. At low temperatures, displacement is a function of mass [18,45] and hence an isotope effect can be expected. An isotope effect on band gap and EXAFS in Ge has been observed [46-48] below Debye temperature and attributed to differences in thermal expansions and electron-phonon interactions for different isotopes. Hence, *an isotope effect on magnetic phase transitions is predicted if $T_c$ is below $\Theta_D$* due to similar effects. In alloy phase transitions, $T_c$ is higher than $\Theta_D$ and high temperature behavior must be considered. At high temperatures, mean-square displacements are independent of mass [45,49]. Hence, an isotope effect in alloy phase transitions is highly unlikely.

Recent experimental results [50,51] where an isotope effect has been found in magnetism confirm all the above conclusions and predictions. A change in the Neel temperature $\Delta T_N/T_N \sim 4\%$ and critical magnetic field $\Delta B_C/B_C \sim 4\%$ is reported for different isotopes in Ref.50. The authors of Ref.50 comment that since the material "*is nonmetallic and so Fermi surface effects have no part in the observed isotope effect*" and attribute the isotope effect to the changes in vibrational amplitudes due to different isotopes. The authors of Ref. 51 state "*We propose that the smaller J' for deuterated NDMAP is caused by the smaller zero-point motion of deuterons leading to less overlap of the electronic wavefunctions in the exchange paths involving hydrogens*". These authors attribute the isotope effect to the differences in vibration amplitudes for different isotopes and clearly suggest that the exchange integral is a function of vibration amplitude and must be written as $J(\overline{u^2})$. This is identical to our analysis above that the DWF component will lead to an isotope effect below $\Theta_D$.

The isotope effect on band structure features has been studied extensively in semiconductors [46-48] in addition to superconductors. Semiconductors are non-metallic as well and as Ref.46-48 show, self energy effects and lattice expansion effects cannot be ignored. Therefore, the conclusion in Ref.50,51 that only the DWF contribution to the isotope effect is significant should be considered to be a preliminary conclusion and more experiments are necessary to separate out the contributions of different phenomena.

The very significant conclusion from the isotope effect observed in Ref.50,51 is that it *cannot be explained by any of the existing models of magnetism such as Ising, Heisenberg, etc. since they do not incorporate the role of thermal vibrations on exchange energies*. The isotope effect observed in Ref.50,51 is due to the same universal phenomena of thermal vibrations that are also responsible for isotope effect in semiconductors [46-48] and therefore universal conclusions can be drawn. Therefore, the most important conclusion that can be drawn from the isotope effect observed in Ref.50 and Ref.51 is that *all theories and models of magnetism must incorporate the role of thermal vibrations on the exchange integral and energies and the exchange integral must be represented as* $J(\overline{u^2})$. Since the vibration amplitude is a function of temperature, the exchange integral can equivalently be represented as *J(T)*.

The above claim is seen more clearly from the following argument. The isotope effect is due to the relatively small differences, $\Delta\left(\overline{u_1^2} - \overline{u_2^2}\right)$, in zero point vibrational amplitudes for different isotopes. It follows that the differences in wavefunctions and exchange integrals, $J(\overline{u_1^2}) - J(\overline{u_2^2})$, will also be relatively small. This relatively small difference in exchange integrals results in different transition temperatures for different isotopes. In contrast, the differences in vibrational amplitudes between two models where one accounts for vibrational amplitude and the other does not as it assumes a static lattice is $\left(\overline{u^2}\right)$ which will be much larger than $\Delta\left(\overline{u_1^2} - \overline{u_2^2}\right)$. Clearly, the difference in wavefunctions and exchange integrals, $J(\overline{u^2}) - J(0)$, between two such models will be much greater than the small difference, $J(\overline{u_1^2}) - J(\overline{u_2^2})$, between two isotopes that leads to an observable isotope effect. Since the vibration amplitude increases monotonically with temperature, the difference, $J(\overline{u^2}) - J(0)$, will also increase with temperature. Hence, the differences in critical temperatures, $\Delta T_c/T_c$ between such models is likely to be much larger compared to $\Delta T_N/T_N$ observed due to different isotopes. To obtain correct results, vibration amplitudes must be incorporated in determining wavefunctions and exchange integrals must be represented as $J(\overline{u^2})$ in all theories and models.

Most of the current models and theories of magnetism are based on the assumption of a static lattice, while in reality atoms in all materials have finite vibration amplitudes at finite temperatures. It is clear that all such models and theories are incorrect and cannot predict the correct $T_N$ or $T_c$. This assumption also leads to contradictions. To conclude that the exchange integral is a function of vibration amplitude when it is small and represent it as $J(\overline{u^2})$ to account for an isotope effect [50,51] but to assume that exchange integral is independent of the much larger vibration amplitude at high temperatures simply because no isotope effect can be observed is a contradiction. The correct conclusion is that the exchange integral must be represented as $J(\overline{u^2})$ at all temperatures. In our view, the implication that <u>exchange integral must be represented as $J(\overline{u^2})$ universally in all theories and models</u> to avoid contradictions is the most significant consequence of the isotope effect observed [50,51] in magnetism. The incorporation of the vibration amplitude or temperature dependence of the exchange integral will lead to new insights into our understanding of magnetism.

The above discussion of the isotope effect reported in Ref.50,51 implies that it is essential to incorporate the role of DWF in electronic structure calculations [30-37] even for magnetic solids which is identical to the conclusion reached by us. In this formalism, wavefunctions and exchange integrals are naturally functions of vibration amplitudes at all temperatures. At low temperatures, since DWF depends on mass, this formalism can readily account for the observed [50,51] isotope effects.

In conclusion, a first principles analysis of order-disorder transition in alloys shows that ordering energy is a function of temperature due to thermal vibrations. The inter-nuclear potential energy term converges if zero point vibrations are incorporated and this method can replace the Ewald sum method. Core energy contributions to the ordering energy are stored exclusively in superlattice lines. The effect of electron-phonon interactions on ordering energy is of the *same order of magnitude* as ordering energy near transition temperatures and cannot be ignored. Ising model and variants are incorrect in explaining alloy phase transitions as they ignore the role of electron-phonon interactions without justification. A theoretical formalism that incorporates the Debye-Waller Factor component of electron-phonon interactions in electronic structure calculations already exists and must be adopted when modeling temperature dependent phenomena. It is suggested that DWF correction will account substantially for the discrepancy between experimental and theoretical ordering energy in $Ni_3V$. Thermal vibrations alter magnetic ordering energy at finite temperatures. The role of electron-phonon interactions in alloy and magnetic phase transitions cannot be ignored and must be incorporated in all models. This will also ensure consistency with x-ray and electron diffraction (alloy transitions) and neutron diffraction (magnetic transitions) results. An isotope effect is predicted for (magnetic) phase transitions if the transition temperature is below Debye temperature. Recent observations of an isotope effect in magnetic phase transitions confirm our above conclusions and imply that the role of electron-phonon interactions must be incorporated in all theories and models of magnetism to avoid contradictions.

We thank an anonymous referee for his comments on the mass independence of DWF at high temperatures and drawing attention to Ref.45. We also thank Prof. P.P. Singh, Dept. of Physics, for his comments and suggestions.

1. W. L. Bragg and E.J. Williams, Proc. Roy. Soc. **A 145**, 699 (1934)
2. F. C. Nix and W. Shockley, Rev. Mod. Phys. **1**, 1 (1938)

3. T. Muto and Y. Takagi, *Solid State Physics – Advances in Research and Applications, F. Seitz and D. Turnbull (Eds),* **1**, 193 (1955)

4. R. Kikuchi, Phys. Rev. **81**, 988 (1951)

5. L. Onsager, Phys. Rev. **65**, 117 (1944)

6. T. R. S. Prasanna, arXiv:cond-mat/0404550

7. F. Ducastelle, *Order and Phase Stability in Alloys* (North-Holland, Amsterdam 1991)

8. D. de Fontaine, *Solid State Physics – Advances in Research and Applications, F. Seitz and D. Turnbull (Eds),* **47**, 33 (1994)

9. A. van de Walle and G. Ceder, Rev. Mod. Phys. **74**, 11 (2002)

10. V. Blum, G. L. W. Hart, M. J. Walorski and A. Zunger, Phys. Rev. B **72**, 165113 (2005)

11. P. Hohenberg and W. Kohn, Phys. Rev. **136**, B864 (1964)

12. W. Kohn and L. J. Sham, Phys. Rev. **140**, A1133 (1965)

13. R.G. Parr and W. Wang, *Density-Functional Theory of Atoms and Molecules,* (Oxford, Oxford 1994)

14. R. F. Stewart and D. Feil, Acta Cryst. **A36**, 503 (1980)

15. V. G. Tsirelson and R. P. Ozerov, *Electron Density and Bonding in Crystals*, (Institute of Physics, Bristol, 1996)

16. P. Coppens, *X-ray Charge Densities and Chemical Bonding*, (IUCr Texts on Crystallography, Oxford, 1997)

17. W. Jones and N. H. March, *Theoretical Solid State Physics, Vol.1*, (Dover, New York, 1985)

18. B.E. Warren, *X-Ray Diffraction*, (Dover, New York, 1990)

19. M. De Graef, *Introduction to Conventional Transmission Electron Microscopy* (Cambridge, Cambridge, 2003)

20. J. Ihm, A. Zunger and M. L. Cohen, J. Phys. C **12**, 4409 (1979)

21. W.E. Pickett, Comp. Phys. Rep. **9**, 115 (1989)

22. J.C. Slater, *Quantum Theory of Molecules and Solids*, vol. 3, (McGraw-Hill, New York, 1967)

23. E. Kaxiras, *Atomic and Electronic Structure of Solids*, (Cambridge, Cambridge 2003)

24. D. R. Chipman and C. Walker, Phys. Rev. B **5**, 3823 (1972)

25. O. Rathmann and J. Als-Nielsen, Phys. Rev. B **9**, 3921 (1974)

26. D. R. Chipman, J. Appl. Phys. **31**, 2012 (1972)

27. N. W. Ashcroft and N. D. Mermin, *Solid State Physics* (Harcourt, New York 1976)

28. J.O. Kephart, B. L. Berman, R.H. Pantell, S. Datz, R.K. Klein and H. Park, Phys. Rev. B **44**, 1992 (1991)

29. G. Buschhorn, E. Diedrich, W. Kufner, M. Rzepka, H. Genz, P. Hoffmann-Stascheck and A. Richter, Phys. Rev. B **55**, 6196 (1997)

30. P.B. Allen and V. Heine, J. Phys. C **9,** 2305 (1976); P.B. Allen and J. C. K. Hui, Z. Phys. B **37**, 33 (1980)

31. P. B. Allen, Phys. Rev. B **18**, 5217 (1978), P. B. Allen and M. Cardona, Phys. Rev. B **23**, 1495 (1981)

32. C. Keffer, T.M. Hayes and A. Bienenstock, Phys. Rev. Lett. **21,**1676 (1968)

33. J. P. Walter, R.R.L. Zucca, M.L. Cohen and Y.R. Shen, Phys. Rev. Lett. **24,** 102 (1970)

34. R.V. Kasowski, Phys. Rev. **187**, 891 (1969)

35. R.V. Kasowski, Phys. Rev. B **8**, 1378 (1973)

36. T. V. Gorkavenko, S. M. Zubkova, V. A. Makara and L. N. Rusina, Semiconductors, **41**, 886 (2007); T. V. Gorkavenko, S. M. Zubkova, and L. N. Rusina, Semiconductors, **41**, 661 (2007)

37. C. Sternemann, T. Buslaps, A. Shukla, P. Suortti, G. Doring and W. Schulke, Phys. Rev. B. **63** 094301 (2001)

38. C. Asker, A. B. Belonoshko, A. S. Mikhaylushkin and I. A. Abrikosov, Phys. Rev. B. **77** 220102(R) (2008)

39. M. Barrachin, A. Finel, R. Caudron, A. Pasturel and A. Francois, Phys. Rev. B. **50** 12980 (1994)

40. O. Libacq, A. Pasturel, D.N. Manh, A. Finel, R. Caudron and M. Barrachin, Phys. Rev. B. **53** 6203 (1996)

41. C. Wolverton and A. Zunger, Phys. Rev. B. **52** 8813 (1995)

42. G. E. Bacon, *Neutron Diffraction* (Clarendon, Oxford 1962)

43. Y. A. Izyumov and R. P. Ozerov, *Magnetic Neutron Diffraction* (Plenum, New York 1970)

44. C. G. Shull, W. A. Strauser and E. O. Wollan, Phys. Rev. **83** 333 (1951)

45. B. T. M. Willis and A. W. Pryor, *Thermal Vibrations in Crystallography* (Cambridge, Cambridge, 1975)

46. C. Parks, A. K. Ramdas, S. Rodriguez, K. M. Itoh and E. E. Haller, Phys. Rev. B **49**, 14244 (1994)

47. M. Cardona and M. L. W. Thewalt, Rev. Mod. Phys. **77,** 1173 (2005)

48. J. Purans, N. D. Afify, G. Dalba, R. Grisenti, S. De Panfilis, A. Kuzmin, V. I. Ozhogin, F. Rocca, A. Sanson, S. I. Tiutiunnikov and P. Fornasini, Phys. Rev. Lett. **100,** 055901 (2008)

49. C. Huiszoon and P. P. M. Groenewegen, Acta Cryst. **A28**, 170 (1972)

50. P. A. Goddard, J. Singleton, C. Maitland, S. J. Blundell, T. Lancaster, P. J. Baker, R. D. McDonald, S. Cox, P. Sengupta, J. L. Manson, K. A. Funk, and J. A. Schlueter, Phys. Rev. B **78**, 052408 (2008)

51. H. Tsujii, Z. Honda, B. Andraka, K. Katsumata and Y. Takano, Phys. Rev. B **71**, 014426 (2005)